\documentclass[twocolumn]{aastex62}

\usepackage{url,graphicx}

\graphicspath{{./figures/}}
\DeclareGraphicsExtensions{.eps}

\newcommand{\adeg}[1]{{#1}$^{\circ}$}
\newcommand{\amin}[1]{{#1}$^\prime$}
\newcommand{\asec}[1]{{#1}$^{\prime\prime}$}
\newcommand{\thour}[1]{{#1}$^{\mathrm{h}}$}
\newcommand{\tmin}[1]{{#1}$^{\mathrm{m}}$}
\newcommand{\tsec}[1]{{#1}$^{\mathrm{s}}$}

\newcommand{\hms}[3]{\thour{#1}\tmin{#2}\tsec{#3}}
\newcommand{\dms}[3]{\adeg{#1}\amin{#2}\asec{#3}}

\newcommand{\chandra}{\emph{Chandra} }

\newcommand{\rosat}{\emph{ROSAT} }

\shorttitle{The Tail of PSR\,J0002+6216}
\shortauthors{Schinzel al.}

\begin{document}

\title{The Tail of PSR\,J0002+6216 and the Supernova Remnant CTB\,1} 

\correspondingauthor{D. A. Frail}
\email{dfrail@nrao.edu}

\author{F.K. Schinzel}
\altaffiliation{An Adjunct Professor at the University of New Mexico.}
\affil{National Radio Astronomy Observatory, P.O. Box O, Socorro, NM 87801, USA}

\author{M. Kerr}
\affil{Space Science Division, U.S. Naval Research Laboratory, Washington, DC 20375, USA}

\author{U. Rau}
\affil{National Radio Astronomy Observatory, P.O. Box O, Socorro, NM 87801, USA}

\author{S. Bhatnagar}
\affil{National Radio Astronomy Observatory, P.O. Box O, Socorro, NM 87801, USA}

\author{D. A. Frail}
\affil{National Radio Astronomy Observatory, P.O. Box O, Socorro, NM 87801, USA}

\begin{abstract}
We have carried out VLA imaging and a {\it Fermi} timing analysis of the 115 ms $\gamma$-ray and radio pulsar PSR\,J0002+6216. We found that the pulsar lies at the apex of a narrowly collimated cometary-like \amin{7} tail of non-thermal radio emission which we identify as a bow-shock pulsar wind nebula. The tail of the nebula points back toward the geometric center of the supernova remnant CTB\,1 (G116.9+0.2) \amin{28} away, at a position angle $\theta_\mu=113^\circ$. We measure a proper motion with 2.9\,$\sigma$ significance from a {\it Fermi} timing analysis giving $\mu$=115$\pm$33 mas yr$^{-1}$ and $\theta_\mu=121^\circ\pm{13}^\circ$, corresponding to a large transverse pulsar velocity of 1100\,km\,s$^{-1}$ at a distance of 2 kpc. This proper motion is of the right magnitude and direction to support the claim that PSR\,J0002+6216 was born from the same supernova that produced CTB\,1. We explore the implications for pulsar birth periods, asymmetric supernova explosions, and mechanisms for pulsar natal kick velocities.
\end{abstract}


\section{Introduction}

Deviations from spherical symmetry appear to be an essential ingredient in successful core collapse supernovae (SN) explosions, and the observational signatures of this asymmetry include the morphology and kinematics of supernova remnant (SNR) ejecta and the natal kick velocities of pulsars \citep[e.g.,][]{hla+17}. The high-velocity outliers in the pulsar velocity ($V_{PSR}$) distribution are of special interest because they provide stringent tests of neutron star kick mechanisms \citep{janka2017}. Characterizing this tail in the kick distribution is a timely topic since, for example, it can affect the fraction of binary neutron star systems that remain bound and their distribution within the host galaxy \citep{berger2014,2018MNRAS.481.4009V}. There are currently only a small number of pulsars that have well enough measured proper motions and distances to robustly claim $V_{PSR}$ in excess of 1000 km s$^{-1}$ \citep{2005MNRAS.360..974H,2017JPhCS.932a2004S}. Similar high velocities have been {\it inferred} from the offset of pulsars from the center of SNRs \citep{fk91,1994ApJ...437..781F} but the burden of proof is high, and in at least two cases follow up scintillation or proper motion measurements do not confirm high velocities \citep{njk96,2008ApJ...674..271Z}.

Recently it was proposed that PSR\,J0002+6216 and the SNR CTB\,1 may be physically associated, based on their angular proximity on the sky and roughly similar distances \citep{zks18}. The 115 ms $\gamma$-ray and radio pulsar PSR\,J0002+6216 is one of the newest additions to the {\it Fermi} sample \citep{cwp+17, wu+18}. It has a large spin-down energy $\dot{{E}}$=1.53 $\times10^{35}$ erg s$^{-1}$ and a dipole magnetic field $B=0.8\times 10^{12}$ G. The SNR CTB\,1 (G116.9+0.2) has been extensively studied at all wavelengths. In optical H$\alpha$ and non-thermal radio it shows a well-defined circular shell of radius $\theta_s$=\amin{17.8} \citep{lrd82}, while the X-rays are centrally concentrated, making it one of a small number of mixed morphology remnants \citep{chp87,ls06,prb+10,knm+18}. Given the angular offset (\amin{28}) and estimates for the age (10$^4$ yr) and distance (2 kpc; see \S\ref{results}), a physical SNR-PSR association would require $V_{PSR}>$1000 km s$^{-1}$. Accordingly, we have carried out new VLA imaging observations and we have re-analyzed {\it Fermi} timing data. Our data support the hypothesis that PSR\,J0002+6216 is a high-velocity pulsar that originated from the same SN that produced the SNR CTB\,1. 

\section{Imaging Observations}\label{obs}

We observed a field toward PSR\,J0002+6216 with the Karl G. Jansky Very Large Array (VLA) as part of project 17B-384 using Director’s Discretionary Time \citep{2009IEEEP..97.1448P}. Observations were carried out on 2017 August 19 with the VLA in the C configuration. We observed in the 1-2\,GHz frequency range (L band) using the standard Wideband Interferometric Digital Architecture (WIDAR) correlator setup for continuum observing with 16 spectral windows, 64x1-MHz wide channels each to get 1 GHz of total bandwidth centered on 1.52 GHz. Data were saved in 5\,s integrations and the total time on source was 63\,min. The radio source J2350+6440 was used as a phase calibrator, while 3C\,48 (J0137+331) was used as both the bandpass and flux density calibrator. 

The correlated visibilities were calibrated using CASA 5.4.0-32 together with the automated VLA pipeline version 41722 \citep{2007ASPC..376..127M,2018AAS...23134214K}\footnote{\url{https://science.nrao.edu/facilities/vla/data-processing/pipeline}}. The pipeline calibrated measurement set had its weights reinitialized prior to imaging according to integration time and bandwidth. The imaging was performed using a customized version of CASA 5.3.0-123 for development of imaging algorithms by NRAO's Algorithms Research and Development group. For imaging, the wideband AWProjection gridding algorithm with conjugate beam models was used in combination with multi-term multi-frequency synthesis (mtmfs) and multiscale clean algorithm \citep{rc11,brg13,rbo16}. The ``Briggs'' weighting scheme (Briggs D., 1995, PhD Thesis, New Mexico Institute of Mining and Technology) was used with a robust setting of 0, corresponding to a compromise between uniform and natural weighting. For w-projection, 128 projection planes were used. The ray-traced primary antenna beam pattern was rotated with 5.0$^\circ$ steps with corresponding parallactic angle. The imaged area covered a region of 3.4$^{\circ}\times$3.4$^\circ$ around the R.A./Dec. pointing of \hms{00}{02}{41.88} and \dms{+62}{18}{2.2}, that is well beyond the half-power point and the first null of the primary beam of the VLA. The rms noise in the final image near PSR\,J0002+6216 was 31 $\mu$Jy beam$^{-1}$ with a synthesized beam of approximately \asec{11}.  This is about a factor of two above the expected theoretical thermal noise limit of 17$\mu$Jy beam$^{-1}$, not taking into account an increase of antenna temperature due to observing in the Galactic plane.
%
%

\section{Timing Analysis}\label{timing}

We constructed a coherent pulse timing model for PSR\,J0002$+$6216 using data from the \textit{Fermi} Large Area Telescope \citep[LAT,][]{Atwood09}.  We selected Pass 8 \citep[P8R3][]{Atwood13,Bruel18} events collected from 2008 August 4 to 2018 November 12 with reconstructed position $<$2$^{\circ}$ from the pulsar position and with reconstructed energy in the interval 100\,MeV to 30\,GeV.  To improve sensitivity, we used a preliminary version of the 4FGL sky model\footnote{\url{https://fermi.gsfc.nasa.gov/ssc/data/access/lat/fl8y/}} to compute the probability that each event originates from the pulsar rather than the background \citep[``photon weights'',][]{Kerr11}.

Because the integration time required to measure the pulse phase (``time of arrival'') is about one month, the time signature of proper motion is only marginally resolved.  Thus we opted to use the \texttt{PINT} pulsar timing package\footnote{\url{https://github.com/nanograv/PINT}} to perform an unbinned analysis.  Moreover, even for large proper motions, the expected timing deviations are small and on the edge of detectability.  
Maximizing the sensitivity demands the sharpest possible model for the pulse profile, but determining the ideal pulse profile is not possible without knowing the timing solution in the first place.  There is thus tension between sensitivity and bias for the proper motion detection, and we take great care to break this circular relationship by adopting the method we describe below.

\begin{figure}
\includegraphics[width=0.45\textwidth]{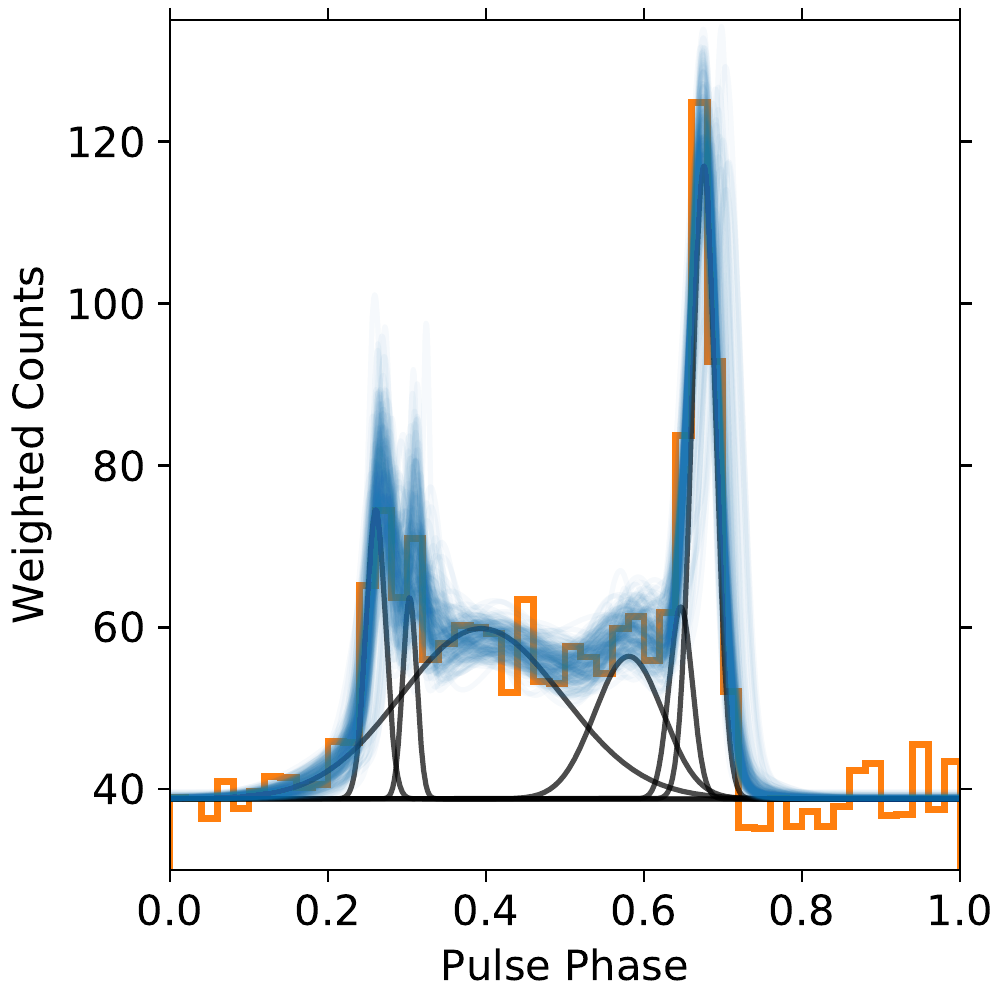}
\vspace{10pt}
\caption{A histogram (orange) of the photon weights using pulse phase from the best-fit timing model (see main text).  Shown as solid black lines are the six wrapped Gaussian components of the analytic profile. The faint blue lines show 200 randomly-chosen realizations of the template from the MCMC fits with the template parameters allowed to vary.}
\label{fig:pulse_profile}
\end{figure}

To determine a good, sharp, starting model of the pulse profile, we optimized both the timing parameters ($\nu$, $\dot{\nu}$, and position) and a series of 12 Fourier components representing timing noise and other unmodelled signals (e.g. proper motion) by maximizing the $Z^2_{10}$ statistic \citep{Buccheri83}, which is sensitive to the power in the first 10 Fourier components of the pulse profile.  We believe this approach allows sufficient freedom in the timing model to produce the sharpest possible profile, but it avoids overfitting by restricting the harmonic content of the pulse profile\footnote{$\gamma$-ray Pulsars with 10 years of \textit{Fermi} data typically require 40 or more Fourier modes to adequately model.}.  To the resulting phases we fit an analytic model, $f(\phi)$, comprising six wrapped Gaussians, shown in in Figure \ref{fig:pulse_profile}.

Next, we studied the timing noise by analyzing a series of models of increasing complexity, specifically including from 2 to 12 Fourier modes with frequencies of 1/T$_{obs}$, 2/T$_{obs}$, etc.  We used \texttt{emcee} \citep{emcee} to explore the posterior distribution of the likelihood $\log\mathcal{L}=\sum_i \log w_i f(\phi_i,\lambda) + (1-w_i)$, with the $w_i$ the photon weights and $\phi_i$ the rotational phases as determined by PINT for a timing model with parameters $\lambda$.  (We adopt uniform priors on all parameters.)  We find that the likelihood improves substantially with the addition of 3 Fourier modes, but that the best-fit power in higher modes drops immediately to a white noise floor.  In tandem, the maximum likelihood only improves marginally with these additional degrees of freedom.  Specifically, we observe a typical increase in the best-fit log likelihood of $\sim$2 for each additional Fourier component.  Because these components satisfy Wilks' Theorem \citep{Wilks38}, such increases are formally insignificant.  Similar conclusions arise from, e.g., the Akaike information criterion \citep{Akaike73}.  Thus, we adopt a three-Fourier component model as our baseline for further analysis.  We note that the overall distribution of power in the Fourier modes is more complicated than the power-law relation seen in many other young pulsars \citep[e.g.,][]{Kerr15}, showing an excess of power on few-year time scales that might originate from an unmodelled glitch recovery preceding \textit{Fermi} observations.

\begin{figure}
\includegraphics[width=0.45\textwidth]{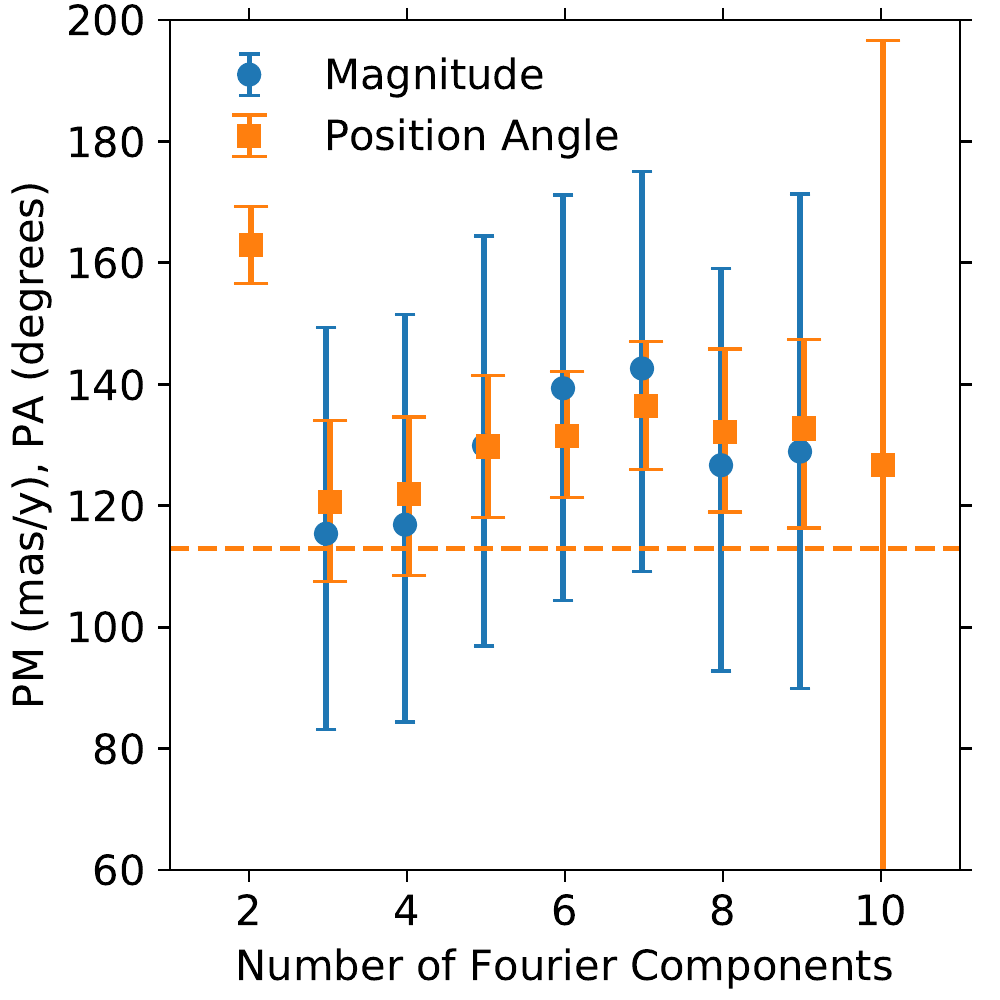}
\vspace{10pt}
\caption{The median magnitude and position angle (measured from North to East) and 1-$\sigma$ confidence range for pulsar timing models with increasing numbers of Fourier components.  The Fourier modes have frequencies of $N/T_{obs}$, with $T_{obs}\sim$10\,yr.  The two-component timing solution produces a poor fit and a discrepant proper motion measurement, while models with four or more Fourier components simply absorb statistical fluctuations and decrease the measurement precision.  With the 10-component model, the proper motion is unconstrained, as there is strong degeneracy between the Fourier component with frequency $\sim$1\,yr and the pulsar position.  Values of the magnitude for 2 and 10 components are above the upper figure boundary.}
\label{fig:pm_waves}
\end{figure}

We next added degrees of freedom for the proper motion to the timing model and studied the posterior over the range of timing noise models.  For our preferred three-component model, the maximum likelihood improves by 5.7, which by Wilks' Theorem has a chance probability of 0.0033, or 2.9$\sigma$ significance.  From the samples of the posterior distribution for each model we computed the 16\%, 50\%, and 84\% quantiles for the magnitude and position angle of the proper motion, and the measurement results are displayed in Figure \ref{fig:pm_waves}.  For our preferred model, we find a total proper motion of 115$\pm$33\,mas\,yr$^{-1}$ at a position angle of 121\degr$\pm$13\degr.  The samples for the coordinate proper motions are correlated ($-0.24$) such that more positive values of $\mu_{\alpha}$ prefer more negative values of $\mu_{\delta}$, leading to a narrower distribution of position angle than the one-dimensional marginalized distributions indicate.  By analyzing the variation of the measurements over the timing noise models, we estimate a systematic uncertainty in the position angle of about 10\degr, while the magnitude is largely unaffected in comparison to the statistical uncertainty.

Finally, we assessed the effect of the assumed pulse profile model on the proper motion measurement by performing Monte Carlo Markov chain (MCMC) runs with the 18 parameters (3 for each Gaussian) of the model allowed to vary, thus marginalizing over these nuisance parameters.  Examples of realizations of the template parameters are shown in Figure \ref{fig:pulse_profile}.  We find similar results: the maximum log likelihood is improved by 5.1, with a chance probability of 0.006 (2.8$\sigma$), and a total proper motion of 112$\pm$39\,mas\,yr$^{-1}$ at a position angle of 125\degr$\pm$19\degr.

These are large values for a pulsar proper motion. In the most recent ATNF Pulsar Catalog \citep{2005AJ....129.1993M}, only 2\% of the sample have measured proper motions as large, and most of these pulsars are at distances $<$1 kpc. We will explore the implications of this result in \S\ref{discussion}.


\begin{deluxetable*}{lr}
\caption{Median and 1-$\sigma$ confidence intervals for \emph{Fermi}-LAT Timing Model\label{tab:timing}}
\tablehead{\colhead{Parameter} & \colhead{Value}}
\startdata
Right Ascension ($\alpha$, J2000)                 & 00$^{\rm h}$02$^{\rm m}$58\fs14(1) \\
Declination ($\delta$, J2000)                     & +62\arcdeg16\arcmin09\farcs52(8) \\
Proper Motion in R.A. ($\mu_\alpha \cos \delta$, mas yr$^{-1}$)  & 97$\pm$ 32\\
Proper Motion in Decl. ($\mu_\delta$, mas yr$^{-1}$)             & -57$\pm$27 \\
Proper Motion Magnitude (mas yr$^{-1}$)           & 115 $\pm$ 33\\
Proper Motion Position Angle (degrees from North) & 121 $\pm$ 13\\
Epoch of position (MJD)                           &  56500.0 \\
Timescale & TDB \\
Solar System Ephemeris & DE421 \\
\enddata
\end{deluxetable*}

\section{Results}\label{results}

\begin{figure*}[htbp!]
\includegraphics[width=7.0in]{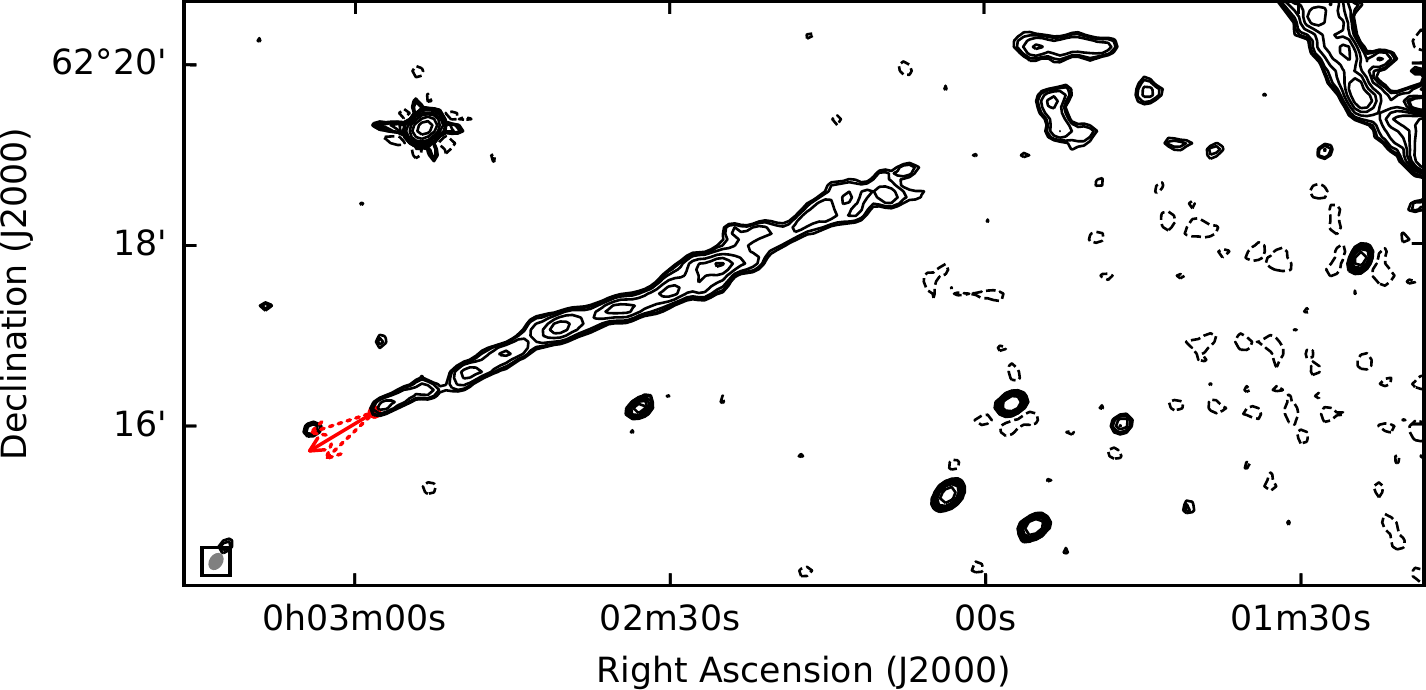}
\caption{Radio continuum image of the cometary tail at 1.5 GHz (20-cm). PSR\,J0002+6216 lies at the base of the arrows at R.A.=\hms{00}{02}{58.17(2)} and Dec.=\dms{+62}{16}{9.4(1)} \citep{cwp+17}. The contour levels are at $-$3, 3, 4, 5, 7, 9, 11, 15, 30, 60, 120 and 240 times the rms noise of 31 $\mu$Jy beam$^{-1}$. The size of the VLA synthesized beam of \asec{12.4}$\times$\asec{9.0} is shown by the ellipse in the bottom left corner. The southeastern shell of the SNR CTB\,1 is visible on the top right hand corner. The solid red arrow shows the future 500 yr proper motion shift at the best fit position angle taken from the posterior distribution of the timing model (Table~\ref{tab:timing}); dashed arrows show 450 yr shifts at the 1$\sigma$ position angle limits.}
\label{fig:pwn}
\end{figure*}

Figure \ref{fig:pwn} shows our 20-cm (1.5 GHz) continuum image toward PSR\,J0002+6216. The pulsar sits at the tip of an elongated cometary tail with Galactic coordinates $l,b$=117.327$^{\circ}$,$-0.074^{\circ}$. The observed emission is dominated by the cometary tail. The  angular extent of this tail is at least \amin{7} and it remains uniformly bright and highly collimated along most of its length. The feature is unresolved (width$<$\asec{13}) for the first half of its length, and is only marginally resolved (width$\sim$\asec{17}-\asec{20}) for the second half. There is extended emission (width$>$\asec{60}) that projects beyond the \amin{7} tail that may be associated with the cometary feature. However,  the rms noise is 20\% higher in this region and thus it is possible that this emission is associated with the nearby SNR CTB\,1 instead.

The total flux density $S_\nu$ of this \amin{7} feature at 1.5 GHz is 14$\pm$0.9 mJy. In order to measure a spectral index, where $S_\nu\propto\nu^\alpha$, we looked for radio images at different frequencies from archival surveys \citep{rtb+97,ijmf17}, and pointed observations of CTB\,1 \citep[e.g.,][]{dw80} but none of these had the requite sensitivity, resolution or field-of-view. We attempted an in-band spectral index measurement by splitting our 1 GHz band into two halves, centered at 1.25 GHz and 1.75 GHz, and imaging and deconvolving separately. The resulting value $\alpha=-0.98\pm0.31$ is not very precise but it is sufficient to suggest that the emission is non-thermal in origin. We imaged the field with full Stokes parameters, but the rms noise is too high to put meaningful limits on the degree of polarization ($<$24\%). 

We also looked for evidence of the tail-like feature at other wavelengths. There is an X-ray point source whose position is consistent with PSR\,J0002+6216 \citep{wu+18,zks18}. The cometary feature is not visible in \rosat images \citep{chp87}, but it lies outside of the field-of-view of deeper pointings made by the narrow-field instruments of {\it ASCA}, \chandra and {\it Suzaku} \citep{ls06,prb+10,knm+18}. Nor is there a structure with this morphology in deep H$\alpha$, [N\,{\sc ii}], [S\,{\sc ii}], or [O\,{\sc iii}] images \citep{fwr+97}.

Based on this morphology and the non-thermal spectral index, we identify G\,117.33$-$0.07 as a sub-class of supersonic or bow-shock pulsar wind nebulae \citep[PWNe;][]{kothes17,2017JPlPh..83e6301K,2017hsn..book.2159S}. These PWNe are typically found after the pulsar has escaped the high pressure environs of their parent SNR. The bow shock is formed as the relativistic pulsar wind is shocked and confined by ram pressure due to the high space velocity of the pulsar through the interstellar medium (ISM). The shocked particles and magnetic energy are swept backward where they emit broadband, non-thermal synchrotron trailing along in a ``tail''.

\begin{figure*}[htbp!]
\includegraphics[width=7.0in]{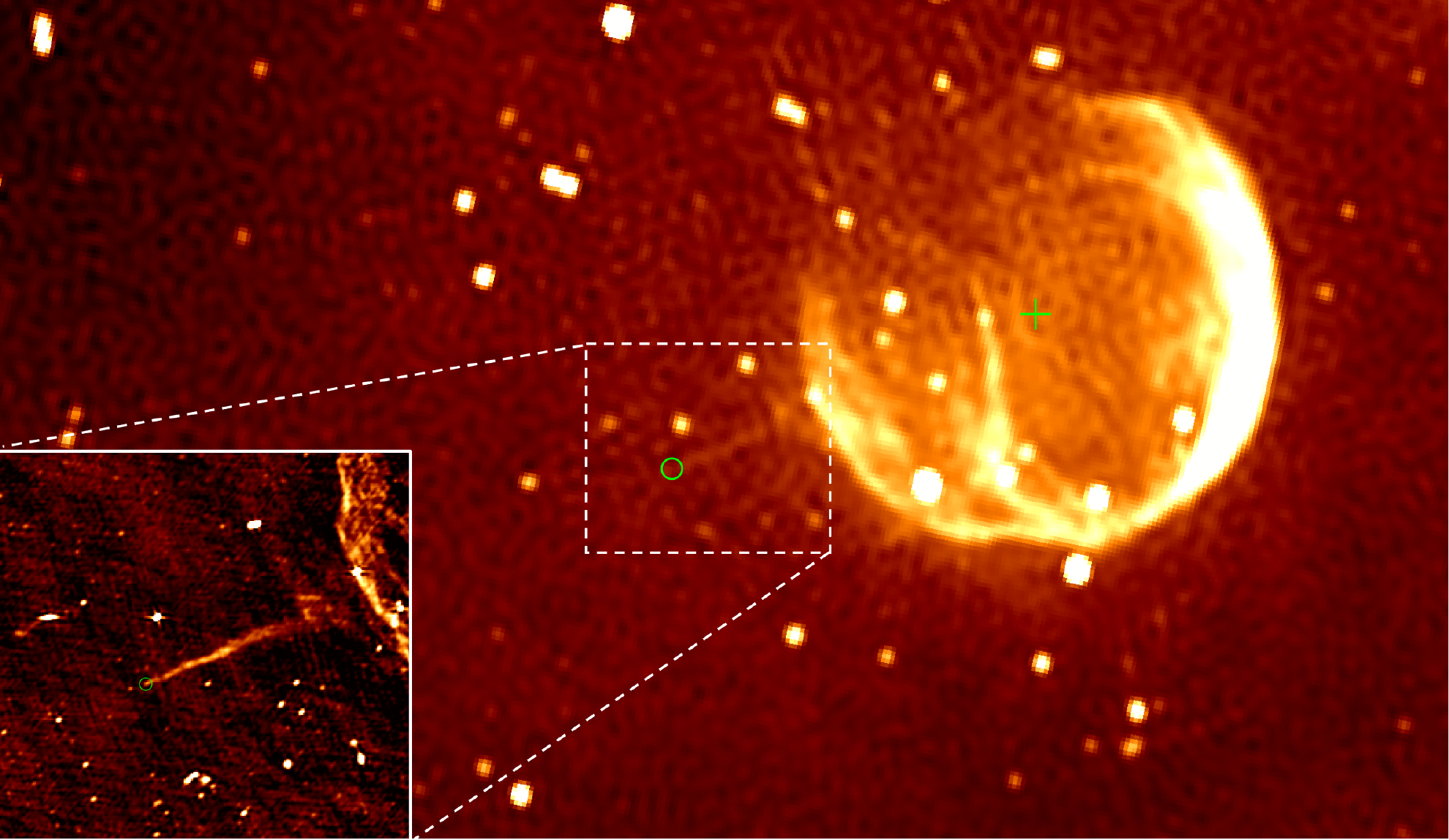}
\vspace{10pt}
\caption{Total intensity image of the SNR CTB\,1 from the Canadian Galactic Plane Survey (CGPS) at 1.42 GHz. False colors start at brightness temperatures of 5.5 K and the maximum is at 8.9 K. The angular resolution and field-of-view are approximately \amin{1} and 1.9$^{\circ}\times$1.1$^\circ$, respectively. A green cross marks the location of the geometric center of the SNR \citep{lrd82} whiles circles indicate the position of PSR\,J0002+6216 \citep{cwp+17}. A faint tail of emission is visible from the PSR to the SNR, pointing back toward the geometric center. The inset is our higher angular resolution 20-cm VLA image of the dashed region taken from Figure \ref{fig:pwn}.}
\label{fig:cgps}
\end{figure*}

The PWN is detected close to the noise threshold in Stokes I images at 1.42 GHz from the Canadian Galactic Plane Survey \citep[CGPS;][]{kffu06}. We show this CGPS image of CTB\,1 in Figure \ref{fig:cgps}. While the resolution is only \amin{1}, the CGPS image includes both single-dish and interferometric data, and thus it faithfully reproduces angular structure on all scales up the resolution limit. The tail in this CGPS image extends from PSR\,J0002+6216 unbroken \amin{11} to the southeastern edge of the SNR CTB\,1. A line from the pulsar along the tail appears to point back to the center of the remnant \amin{28}$\pm$\amin{1} away, suggesting a possible common origin. To test this hypothesis we measured the positions of 10 radio peaks along the \amin{7} feature and we fit a linear least squares solution to a line of these data. The extrapolation of this \amin{7} line passes within \asec{5} and \asec{11} of the geometric center of CTB\,1 {\it independently} derived by \citet[$\alpha_{1950}=23^{\rm h}45^{\rm m}45\fs$, $\delta_{1950}=62^\circ10.5'$; ][]{lrd82} and \citet[$\alpha_{2000}=23^{\rm h}59^{\rm m}16\fs$, $\delta_{2000}=62^\circ 27'$; ][]{kffu06}, respectively. This offset is considerably smaller than the $\pm$\amin{1} position uncertainty of the geometric center of CTB\,1. Note that we are assuming here that the geometric center is a good proxy for the origin of the supernova event, which is not necessarily the case \citep{gva04}.

A further test of the association is to look for proper motion of the pulsar. The age of CTB\,1, derived from X-rays, assuming Sedov evolution, is 9-13.3 kyr \citep{chp87,ls06}, while similar values of 7.5 kyr \citep{hc94} are obtained from optical data. We adopt a value of 10 kyr, recognizing that this value is uncertain by about 20\%. With the \amin{28} PSR-SNR offset\footnote{\citet{zks18} use a PSR-SNR angular offset of \amin{17} rather than \amin{28}, apparently in error.} we predict the proper motion of PSR\,J0002+6216 to be 168$\pm$35\,mas\,yr$^{-1}$ in the southeast direction with a position angle $\theta_\mu=113^\circ$ (measured from north through east). This is in good agreement with pulsar timing proper motion measurement of $\mu=115\pm33$\,mas\,yr$^{-1}$ at a position angle $\theta_\mu=121\pm{13}^\circ$. Moreover, we can rule out a proper motion with magnitude $<$63 mas yr$^{-1}$ with 95\% confidence.  If we constrain the proper motion position angle to be $113\pm3^{\circ}$, we find $\mu=121\pm30$\,mas\,yr$^{-1}$ and the 95\% confidence range increases to $>$73 mas yr$^{-1}$.  We conclude that PSR\,J0002$+$6216 passed through the center of CTB\,1 between 10 and 20 kyr ago.



\section{Discussion\label{discussion}}

In \S\ref{results} we have shown that PSR\,J0002+6216 lies at the ``head'' of a cometary-shaped feature whose tail points back to the geometric center of the SNR CTB\,1. We argue that this feature is a bow-shock PWN shaped by the pulsar's supersonic motion through the ISM, and that the PSR and SNR may have a common origin. From a pulsar timing analysis we measure a proper motion that agrees in magnitude and direction to the predicted value derived from the PSR-SNR angular offset and the SNR age. In the following subsections we will explore the implications of each of these main results.

\subsection{The velocity and distance to PSR\,J0002+6216\label{zoom}}

Our measured proper motion from the timing analysis (\S\ref{timing}) of $\mu=115\pm$ 33 mas yr$^{-1}$ corresponds to a transverse pulsar velocity $V_{PSR}$=550$\times{d_{kpc}}$ km\,s $^{-1}$, where $d_{ kpc}$ is the pulsar distance in kpc. The distance to  PSR\,J0002+6216 can be estimated from its dispersion measure DM=218.6 pc cm$^{-3}$ \citep{wu+18}. The two models for the Galactic distribution of ionized gas give $d$=7.9 kpc \citep{cl02} and $d$=6.4 kpc \citep{ymw17}, respectively, leading to a suspiciously large $V_{PSR}$=3500-4300 km\,s $^{-1}$. Both \citet{wu+18} and \citet{zks18} have also noted that the DM distance would also require that the $\gamma$-ray efficiency of PSR\,J0002+6216 exceed unity. They derive a pulsar distance from an empirical relationship between the $\gamma$-ray luminosity and $\dot{{E}}$ of 2.0 kpc and 2.3 kpc, respectively. At $d$=2.3 kpc our measured proper motion gives $V_{PSR}$=1260$\pm$ 360 km\,s $^{-1}$. While still large, there are other pulsars with well-measured proper motions with comparable values of $V_{PSR}$ including PSRs B1727$-$47, B2011+38, and B2224+65 \citep{2005MNRAS.360..974H,2017JPhCS.932a2004S}. 

The foregoing suggests that the DM-based distance to PSR\,J0002+6216 may be an overestimate and that there is an excess source of free electrons beyond that assumed in the Galactic models \citep{cl02,ymw17}. Following \citet{2014ApJ...797...70K} we can attempt to derive the size, $L$, and distance, $d$, to this excess contribution $DM_e$ using three observational constraints: the diffuse H$\alpha$ emission, the known ionizing stars, and the DM distribution of local pulsars. At an angular resolution of 1$^\circ$ the Wisconsin H$\alpha$ Mapper Northern Sky Survey \cite[WHAM;][]{2003ApJS..149..405H} measures an integrated H$\alpha$ intensity of 28.5 Rayleigh, or EM=65 pc cm$^{-6}$, where we have used the conversion of 1 Rayleigh corresponding to EM=2.25 pc cm$^{-6}$ for an ionized gas with electron temperature of 8000 K. We can obtain a second estimate of EM on arcminute scales from the deep H$\alpha$ image of \citet{fwr+97}. Along the southeastern edge of the nearby CTB\,1 they measure an extinction-corrected H$\alpha$ surface brightness 6.6$\times 10^{-16}$ erg cm$^{-2}$ s$^{-1}$ arcsec$^{-1}$ or 2.8$\times 10^{-5}$ erg cm$^{-2}$ s$^{-1}$ sr$^{-1}$, or EM=260 pc cm$^{-6}$, where we have used the conversion of 1 Rayleigh=2.41$\times 10^{-7}$ erg cm$^{-2}$ s$^{-1}$ sr$^{-1}$ \citep{2014ApJ...797...70K}. From the same deep H$\alpha$ images we estimate that the diffuse H$\alpha$ emission toward PSR\,J0002+6216 is about 10$\times$ fainter, or EM=30 pc cm$^{-6}$. Thus any nebular source of excess electrons is constrained to have EM$\leq$30-65 pc cm$^{-6}$, where EM=DM$_e^2$/$L_{pc}$ and $DM_e$=$n_e\times$$L_{pc}$.
 
 It is straightforward to account for this excess contribution if the pulsar lies at a distance $d\gtrsim$3.4 kpc. At that distance the excess DM required in the \citet{ymw17} model over the observed DM for PSR\,J0002+6216 is DM$_e\simeq$50 pc cm$^{-3}$ and EM=2500/$L_{pc}$. $DM_e$ is about a factor of two higher for the \citet{cl02} model. Neither of these models account for the fact that PSR\,J0002+6216 passes within \amin{8} of the line of sight of HD\,225160, an 8th magnitude O8 blue supergiant with a parallax distance from {\em Gaia} DR2 of 3.4$\pm$0.4 kpc \citep{2018A&A...616A...1G}.  Such a star will ionize an extended region and be surrounded by a Str\"omgren sphere whose size $R_s$ for a uniform density $n_e$ is given by $R_s$=$n_{e}^{-2/3}U$. The excitation parameter $U$ (in units of pc cm$^{-2}$) is calculated from stellar atmospheric models and conveniently expresses the ionizing flux as a function of spectral type and spectral class \cite[e.g.,][]{1973AJ.....78..929P}. A line-of-sight that intersects an H\,{\sc ii} region with impact parameter $R$ will see an excess dispersion measure $DM_e$=2$R_s{n}_e\sqrt{1-({R}/{ R}_s)^2}$ where $L=2\sqrt{{R}_s^2-{R}^2}$ \citep{1969MNRAS.146..423P}. For ISM number densities $n_e<30$ cm$^{-3}$ the line-of-sight of the pulsar ($R$=8 pc, i.e., \amin{8} at 3.4 kpc) always passes through the Str\"omgren sphere of HD\,225160.  The derived value of $DM_e$ is relatively insensitive to n$_e$ with $DM_e$=70-150 pc cm$^{-3}$ for n$_e$=0.1 to 1.0 cm$^{-3}$ and the corresponding emission measures are 10-150 pc cm$^{-6}$. Thus an H\,{\sc ii} region ionized by HD\,225160 can easily explain the required $DM_e$ and is consistent with the EM constraints from H$\alpha$.
 
At smaller distances a third constraint comes into play since the angular size of the nebula providing $DM_e$ cannot be so large as to affect the DMs of known pulsars in the vicinity. Within a 6$^\circ$ radius of PSR\,J0002+6216 there are 12 other pulsars. The majority (9) have DMs in the range of 100-125 pc cm$^{-3}$ with an average distance of 2.4 kpc, while the remaining have DMs of about 200 pc cm$^{-3}$. PSR\,B2351+61 with DM=94 pc cm$^{-3}$ is closest on the sky to PSR\,J0002+6216, which at 1.1$^\circ$ provides a constraint on $L/d$. For example, at $d$=2.3 kpc the maximum dimension the nebular region can have in the plane of the sky is $L$=44 pc. At this distance the \citet{ymw17} model predicts DM$_e\simeq$130 pc cm$^{-3}$ for PSR\,J0002+6216 and EM=1.7$\times{10^4}/L$. To satisfy the deep H$\alpha$ EM constraints (EM$\leq$30-65 pc cm$^{-6}$) any putative nebula would need to be significantly elongated (10:1) or there must be large extinction toward  PSR\,J0002+6216, reducing the observed EM. We examined 3D extinction models in this direction \citep[e.g.,][]{sale14}. Within \amin{12} of PSR\,J0002+6216 the extinction (defined at $\lambda$=549.5 nm) has a value A$_{\circ}$=1.3 mag by $d$=2.3 kpc and nearly doubles by $d$=3.5 kpc. The distribution of dust is quite patchy; within a 1$^\circ$ radius $A_{\circ}$ at 1 kpc varies from 0.43 mag to 1.38 mag, while at 3 kpc the extinction range is 1.2 to 2.4 mag. This is consistent with the observed column density and optical extinction toward the nearby SNR CTB\,1, varying from $A_{V}$=1 to 2 mag \citep{fwr+97,knm+18}. It becomes exceedingly difficult, however, to satisfy all constraints at smaller distances. At $d$=1 kpc, for example, PSR\,J0002+6216 is less an outlier in the pulsar velocity distribution, but the local DM constraints give $L\leq${20} pc and the \citet{ymw17} model predicts DM$_e\simeq$200 pc cm$^{-3}$ and EM=4$\times{10^4}/L$. To satisfy the EM constraints would require a narrow tube of nearly 1000 pc in length and width 20 pc, a highly unlikely configuration. 

Summarizing, we find that the DM-distance gives unreasonable values for $V_{PSR}$ and requires that the pulsar $\gamma$-ray efficiency exceeds unity. We argue that there is evidence that there are additional source of free electrons along the line of sight. The nearest distance that is consistent with the existing observational constraints is $d$=2 kpc. This is the same distance derived from an empirical relation between between the $\gamma$-ray luminosity and $\dot{{E}}$. For the remaining discussion and derivations we will adopt $d$=2 kpc ($d_2\equiv$1) for PSR\,J0002+6216, recognizing that this estimate is uncertain.

\subsection{The PWN of PSR\,J0002+6216\label{pwn}}

Within the sub-class of bow-shock or supersonic PWNe there are a diverse range of morphologies \citep[see list in][]{2017JPlPh..83e6301K}, likely driven by small-scale variations in the density and magnetic field of the ISM, and/or the different orientations of the pulsar magnetospheric spin axis with respect to the velocity vector \citep{2014AN....335..234B,bl18,trl18}. The PWN that most resembles G\,117.33$-$0.07 is G\,315.78$-$0.23 around PSR\,J1437$-$5959, a moderately aged pulsar with a period of 61.7 ms and a spin-down energy $\dot{{E}}$=1.5 $\times10^{36}$ erg s$^{-1}$ \citep{cng+09,nbg+12}. Both PWNe have the same angular extent with narrow heads and a narrow, uniformly bright tail. Here we will follow \citet{nbg+12} in calculating some standard PWN properties for G\,117.33$-$0.07 and compare them to G\,315.78$-$0.23. Our estimates should be viewed as preliminary since we have only a single radio image (1-2 GHz) in which G\,117.33$-$0.07 is only marginally resolved (Figure \ref{fig:pwn}) and the synchrotron spectrum is poorly constrained (\S\ref{results}).

The size of the bow shock region $r_s$ can be estimated by equating the ram pressure of the fast-moving pulsar $\rho{V}^{2}_{PSR}$ to the wind pressure of the pulsar $\dot{{E}}$/4$\pi r^2_s$c \citep[see][]{fggd96}. Since the source is unresolved we assume an isotropic wind with $\dot{{E}}$=1.53 $\times10^{35}$ erg s$^{-1}$ for PSR\,J0002+6216 and adopt ISM number densities of 0.1-1 cm$^{-3}$, which bracket the density range derived from HI and X-ray observations \citep[e.g.,][]{yuk04,ls06}. We have assumed a 10\% He mass in the ISM when calculating $\rho$. From this we derive $r_s$=(1-4)$\times 10^{-3}d_2^{-1}$ pc. This is identical to the small standoff distance of $r_s\simeq2.4\times 10^{-3}$ pc derived for G\,315.78$-$0.23, but since G\,117.33$-$0.07 is 4x closer it might be possible to resolve its bow shock (\asec{0.1-0.4}) with future radio observations. An alternative way to express this pressure balance is to write it in terms of the ISM pressure \citep{2008ApJ...684..542K}, which is generally more robustly determined than the ISM density. Now $\rho{V}^{2}_{PSR}$ is re-written as $\gamma{\mathcal{M}^2}P_{ISM}$, where $\gamma\equiv5/3$ is the ISM adiabatic index, $\mathcal{M}$ is the Mach number, and $P_{ISM}$ is a typical thermal ISM pressure of 10$^{-12}$ dyn cm$^{-2}$. The estimated shock Mach numbers are $\mathcal{M}\simeq{200}$ for both G\,315.78$-$0.23, and G\,117.33$-$0.07. These large Mach numbers and the narrowness of the PWN tails are suggestive of a high pulsar velocity in both systems.

While the head of G\,117.33$-$0.07 is unresolved with our \asec{12} beam, the linear dimensions of the \amin{7} tail are $l$=4.1$d_2$ pc and the half-width of the end of the tail is $h$=0.09$d_2$ pc. This implies that the PWN has expanded by about a factor of 45 from the tip of the head (r$_s$) to the end of the tail ($h$). At $d$=8 kpc the corresponding dimensions of G\,315.78$-$0.23 are a tail length of $l$=20 pc and half-width of $h$=1.2 pc.  Provided that the distance of 8 kpc is correct, G\,315.78$-$0.23 has the longest radio tail among the known bow-shock PWNe, while G\,117.33$-$0.07 has an average tail length \citep{2008ApJ...684..542K,nbg+12,2017JPlPh..83e6301K}. We approximate the volume of G\,117.33$-$0.07 using a half-cone of radius $r_s$ plus a frustum of a right circular cone with length $l$ and radii $r_s$ and $h$, giving a volume $V=10^{54}d_2^3$ cm$^{-3}$.

The slope of the radio spectrum of G\,117.33$-$0.07 of $\alpha=-0.98\pm0.31$ (\S\ref{results}), is atypical of radio PWNe and is more similar to the slopes of X-ray PWNe \citep{kothes17, 2017hsn..book.2159S}. We caution however that $\alpha$ was estimated within a narrow 1 to 2 GHz band and needs to be better measured over a wider frequency range. For now we will calculate PWN parameters with both $\alpha=-1$ and a more typical radio PWN of $\alpha=-0.3$. The radio luminosity $L_{R}$ is obtained by integrating from 10 MHz to 100 GHz, giving values of 2.7$\times{10}^{30}d_2^2$ erg s$^{-1}$ ($\alpha=-0.3$) and 9.3$\times{10}^{29}d_2^2$  erg s$^{-1}$ ($\alpha=-1$). The radio efficiency $L_{R}$/$\dot{{E}}$ is $\sim$10$^{-5}$, typical of other radio PWNe \citep{1997ApJ...480..364F,2000MNRAS.318...58G}. 

The synchrotron spectrum can be used to derive useful PWN parameters such as the strength and energy density of the magnetic field, the energy density of the relativistic particles, and the lifetime of the radiating electrons. The standard expressions are given in \citet{1970ranp.book.....P} but we use equations 8-10 from \citet{nbg+12} in which the magnetic field B, the synchrotron lifetime t$_{syn}$ and the relativistic gas and field pressure in the tail $P_{tail}$ are conveniently expressed in terms of the lesser-known wind parameters such as the ion to electron energy density ratio $\eta$ and magnetic to particle (electrons and ions) energy density ratio $k_m$. If we adopt their values of $\eta$=0 (i.e. no ions) and $k_m$=0.1 (electron-dominated PWN) we derive $B$=30$d_2^{-2/7}$ $\mu$G, $t_{syn}$=1$d_2^{3/7}$ Myr (at 10 GHz), and $P_{tail}$=1.6$\times{10}^{-10}{d_2^{-4/7}}$ dyn cm$^{-2}$. Remarkably, apart from a distance scaling, these values for G\,117.33$-$0.07 are identical to the values derived for G\,315.78$-$0.23. We have integrated the synchrotron spectrum in the range 10$^7$ Hz to 10$^{13}$ Hz, but the values are relatively insensitive to the outer frequency range and they are not sensitive to which value of $\alpha$ we use. 

G\,117.33$-$0.07 and G\,315.78$-$0.23 appear to be near morphological twins of each other, although the tail of G\,117.33$-$0.07 is five times smaller at our adopted distance of 2 kpc. Both PWNe are powered by high-energy pulsars moving at high velocity away from a parent SNR (see \S\ref{ass}). Their PWNe are characterized by a small shock standoff distance and a large Mach number ($\mathcal{M}\simeq200$) with long synchrotron-emitting tails which are likely dominated by relativistic particles rather than B-fields. In the case of G\,315.78$-$0.23 the magnetic field is detected and is seen to be aligned with the tail. For G\,117.33$-$0.07 we lack such data. Future broadband radio continuum and polarimetric observations will fully resolve the tail and measure the magnetic field orientation, and together with X-rays will better constrain the spectrum and thus the parameters derived above. Likewise, higher angular resolution observations of the head of G\,117.33$-$0.07 would also show whether the PWN morphology is being shaped by the geometry of the pulsar magnetopshere and/or the pulsar proper motion. 

The age of the radio PWN is obtained simply by dividing its angular size with the pulsar proper motion giving $t_{R}$=3600 yr. As expected, $t_{R}$ is much smaller than the 1 Myr radiative lifetime of the radio-emitting electrons but at X-ray energies (assuming B=30 $\mu$G from above) $t_{syn}\simeq$200 yr (1 keV). Thus we might expect to see a smaller X-ray PWN scaled approximately by the ratio of \amin{7}($t_{syn}$/$t_{R})\simeq$\asec{20}, or even longer if the streaming velocity of the post-shock wind is a significant fraction of the speed of light. The prospects for the detection of an X-ray PWN appear good since for typical X-ray PWN efficiencies of 10$^{-3}$ the X-ray luminosity $L_X\simeq{10}^{32}$ erg s$^{-1}$ \citep{2008ApJ...684..542K}, and G\,117.33$-$0.07 is relatively nearby at 2 kpc with only modest gas column densities expected $\sim$10$^{21}$ cm$^{-2}$. 

\subsection{The PSR\,J0002+6216 SNR CTB\,1 association\label{ass}}

All neutron stars, given a substantial kick at the time of birth, will eventually escape their parent SNR on a timescale $\tau_{esc}=({E}_\circ/{n}_\circ)^{1/3}$\,$V^{-5/3}_{PSR}$ that is only weakly dependent on the explosion kinetic energy E$_\circ$ and the ISM density n$_\circ$, but is sensitive to the magnitude $V_{PSR}$ \citep{vag+03,bap+17}. Our measured proper motion of PSR\,J0002+6216 is of the right magnitude and direction to support the claim that it was born from the same SN that produced the SNR CTB\,1 \citep{zks18}. Likewise, the direction and the morphology of the 7-\amin{11} (4.1-6.4 pc) tail of the PWN suggest a physical connection between a high-velocity pulsar and its SNR. While promising, a secure PSR-SNR association also requires good agreement of distances and ages \citep{kaspi96}.

The distance to CTB\,1 can be reliably estimated since it is claimed to be part of a much large star-forming complex in the Perseus arm \citep{fich86}. On smaller scales CTB\,1 is embedded in a neutral hydrogen (H\,{\sc i}) hole and is interacting with neutral gas along its bright radio continuum edges at a local standard of rest velocity of $-30$ km s$^{-1}$ \citep{lrd82}. This value agrees remarkably well with the mean velocity derived from optical spectroscopy \citep[][and references therein]{hc94}. Thus we adopt the kinematic distance of 2.0$\pm$0.4 kpc from \citet{lrd82}, but we note that \citet{yuk04} find the same H\,{\sc i} structures and velocities but argue instead that CTB\,1 is a blue-shifted Local arm object at a distance of 1.6 kpc. Larger kinematic distances quoted for CTB\,1 use older IAU Galactic rotation parameters \citep{2014ApJ...783..130R,2018ApJ...856...52W}, or neglect to account for the well-known non-circular motions toward the Perseus arm in this direction \citep{2014ApJ...790...99C,2018IAUS..336..168S}.

In \S\ref{zoom} we argued that the nominal DM distance to PSR\,J0002+6216 was an overestimate and we showed that there were additional sources of free electrons along the line of sight. The nearest distance that PSR\,J0002+6216 could be without violating several observational constraints is $d$=2 kpc, although we can not rule out $d\simeq$3.5 kpc. If PSR\,J0002+6216 lies at 2 kpc its distance would agree with the kinematic distance of SNR CTB\,1. We note that this is the same distance as the adjacent PSR\,B2334+61 (DM=58.41 pc cm$^{-3}$) SNR G\,114.3+0.3 association \citep{kpha93} 3$^\circ$ away, and it would place both PSR-SNR associations in the same star-forming complex within the Perseus arm.

The characteristic age of the pulsar $\tau_c$=306 kyr greatly exceeds the SNR age of $\sim$10 kyr by a factor of 30. If $\tau_c$ is the correct age, our proper motion value would suggest that the pulsar was born far away (11$^\circ$) and that the PWN and CTB\,1 are a chance alignment. \citet{nbg+12} argue that the chance alignment of a PWN tail pointing back to the geometric center of an SNR is exceedingly unlikely. These two ages could be reconciled if PSR\,J0002+6216 was born with an initial period (P$_\circ$) close to its current period P=115 ms \citep{ctk94,zks18}. If so, PSR\,J0002+6216  would be one of a growing number of young PSRs in SNRs \citep[including PSR\,J1437$-$5959;][]{cng+09} for which P$_\circ$/P $\not\ll$1 \citep{pt12}, casting doubt on the reliability of $\tau_c$ for assessing PSR-SNR associations. 

Comparing independently derived pulsar velocities provides another consistency check on the association. At a distance of 2$\pm{0.4}$ kpc and age 10$\pm{0.2}$ kyrs, the \amin{28}$\pm$\amin{1} PSR-SNR offset predicts $V_{PSR}$=1600$\pm$450 km\,s $^{-1}$, where the uncertainty is calculated from the quadrature sum of the errors on age, distance and angular offset. From our measured proper motion and the 2 kpc distance we derive an age-independent $V_{PSR}$=1100$\pm$315 km\,s$^{-1}$. A third method for estimating the transverse velocity of the pulsar is given by \citet{fggd96} where the velocity of the pulsar can be written as $V_{PSR}$=$V_s\,\beta$/$c_\circ$, where $V_s$ is the shock velocity of the SNR, $\beta$ is the fractional offset of the pulsar from the center of the remnant normalized by the remnant radius, and $c_\circ$ is a constant equal to 2/5 for a remnant in the Sedov phase. This method has the advantage of being independent of both the age and the distance of the association, but it does weakly depend on the evolutionary state ($c_\circ$) of the remnant. A lower bound of $V_s\gg$ 100 km s$^{-1}$ comes from optical line ratios such as [O\,{\sc iii}]/H$\beta$ \citep{fwr+97}. Thermal plasma is seen from CTB\,1 with X-ray temperatures of 0.2 keV to 0.28 keV, which corresponds to V$_s$=410-480 km s$^{-1}$ \citep[see eqn. 2 of][]{ls06}, and thus $V_{PSR}$=1780$\pm$150 km\,s $^{-1}$.  The consistency of these three $V_{PSR}$ estimates does bolster the claim of a PSR-SNR association.

\subsection{PSR\,J0002+6216 and pulsar kick mechanisms\label{kick}}

One of the more robust conclusions from this work is that PSR\,J0002+6216 is a high-velocity pulsar. Most pulsars have average transverse velocities of order 250 km s$^{-1}$ \citep{2005MNRAS.360..974H,vi2017}, so PSR\,J0002+6216 is a rare outlier with $V_{PSR}>$1000 km s$^{-1}$. While it has long been known that pulsars receive a substantial kick at birth, the debate about the mechanism(s) is still an active research topic. Numerous kick mechanisms have been proposed including binary disruption, asymmetric neutrino emission, jets that accelerate the pulsar, and hydrodynamic instabilities \citep{1996ApJ...456..738I, 2001ApJ...549.1111L, 2006A&A...457..963S,2007ApJ...660.1357N, 2013A&A...552A.126W,2018arXiv181008620K}. A high-velocity pulsar like PSR\,J0002+6216 poses a strong challenge to these models. From the magnitude of the velocity alone we can rule out binary disruption  \citep{1996ApJ...456..738I,1998ApJ...496..333F}. Some jet models predict large scale morphological distortions in the SNR along the axis defined by the pulsar's proper motion \citep{2018ApJ...855...82B}. This is not seen in the case of SNR CTB\,1 which is remarkably circularly symmetric in both optical H$\alpha$ and non-thermal radio.  
 
A general prediction for several natal kick models, that has some observational support \citep{2007MNRAS.381.1625J,2007ApJ...660.1357N}, is an alignment between the pulsar's rotation axis and the direction of the pulsar velocity \citep{sp98}. \citet{wu+18} have fit the pulse profiles of PSR\,J0002+6216 to constrain some of the magnetospheric parameters using outer gap (OG) and slot gap emission models (TPC).  They measure the angle between the rotation axis and the line of sight $\xi_{TPC}=54^\circ\pm2\degr$ and $\xi_{OG}=58\degr^{+25}_{-1}$, and the angle between the  rotation axis and the magnetic axis $\alpha_{PC}=64\degr^{+3}_{-2}$ and $\alpha_{OG}=69\degr^{+8}_{-1}$. These estimates are highly model dependent and are subject to systematic errors of 10$^\circ$ or larger, so that no strong conclusions should be drawn from these values.

The angle $\xi$ is particularly interesting since for those PWNe with a toroidal wind it describes the angle that the axis of the torus makes with respect to the plane of the sky \citep{2001ApJ...556..380H,2004ApJ...601..479N}, and it can have a strong impact on the morphology of bow-shock PWNe \citep{bl18}. Unfortunately, while we know $\xi$, we do not know its position angle on the plane of the sky to compare with $\theta_\mu$, the direction of the pulsar's motion. This angle is derived from polarization measurements of the radio pulse profile and PSR\,J0002+6216 is too faint to make such measurements in the near term. However, if the velocity and rotation axis are aligned then we can say that the 3D velocity would be $(\cos\xi)^{-1}\simeq1.7\times$ larger than the transverse value $V_{PSR}$. Another trend expected from these finite duration kick models is that the fastest moving pulsars should have the longest birth periods \citep[Figure 12 of][]{2007ApJ...660.1357N}. Interestingly for our high-velocity PSR\,J0002+6216, we agree with \citet{zks18} that if the PSR-SNR association is real, the pulsar birth period P$_\circ\simeq$P=115 ms (\S\ref{ass}).

Recent work on young ejecta-dominated SNRs has established a link between pulsar kick velocities and SNR asymmetries, as measured via X-ray morphologies \citep{kmj+18}. These results support the hydrodynamic instability kick model in which the SN recoil expels the pulsar in one direction and the newly-synthesized heavy elements from the core in the opposite direction \citep{janka2017}. The high velocity of PSR\,J0002+6216 could pose a particularly strong test of this mechanism. To illustrate this we note that the kinetic energy of a 1.5 M$_\odot$ neutron star moving with $V_{PSR}$=1100 km s$^{-1}$ is $E_k=0.2\times{10}^{50}$ erg. This is likely a lower limit since we have only the transverse 2D velocity of the pulsar. Several different analyses have all concluded that CTB\,1 is a low-energy SNR with an explosion kinetic energy $E_\circ\simeq{10}^{50}$ erg \citep{chp87,fwr+97,ls06}. As $E_k$ is a significant fraction of E$_\circ$ it may be difficult for a low-energy SN like CTB\,1 to impart a substantial natal kick to PSR\,J0002+6216, but if the mechanism does work it would predict strong asymmetries.

Is there a signature from the natal kick given to PSR\,J0002+6216 in either the large-scale X-ray morphology of CTB\,1 or the distribution of its ejecta? The X-ray emission from CTB\,1 is concentrated within the radio shell with its X-ray centroid shifted to the northwest \citep[e.g.,][]{prb+10}. This asymmetry could be due to the natal recoil, but since CTB\,1 is a middle-aged SNR, it is more likely that the morphology is dominated by its interaction with the ISM \citep{ls06}. There is ample evidence from the H\,{\sc i} kinematics that CTB\,1 is strongly interacting with the surrounding gas \citep{lrd82}, and from X-ray, H$\alpha$ and H\,{\sc i} velocity images we see that the northeast rim of the remnant has broken out of this cavity and is expanding into a lower density region \citep{chp87,fwr+97,yuk04}.  Thus we find no evidence to suggest that a kick signature is seen in the morphology of CTB\,1. 

Despite CTB\,1 being a mixed morphology remnant that is strongly shaped by the ISM, the analysis of X-ray spectra shows evidence for enhanced heavy element abundances \citep{ls06}. The abundance ratios of these ejecta are consistent with CTB\,1 being an oxygen-rich SNR produced in a core-collapse SN with a progenitor mass of 13-15 M$_\odot$ \citep{prb+10}. In a deep (82 ks), narrow-field {\it Suzaku} pointing of the center of CTB\,1, \citet{knm+18} carried out a spectral analysis of three regions, finding enhanced iron abundances in one region (region E) that they interpret as originating from asymmetric SN ejecta. This region encompasses a large solid angle of the SNR, so it is not possible to say at this time if the centroid of this ejecta asymmetry is directed away from the direction of motion of PSR\,J0002+6216. Future searches for heavy element asymmetries should be made along an axis defined by the PWN tail and the geometric center of the SNR.

\section{Summary and Conclusions\label{summary}}

In this paper we have presented an analysis of VLA 20-cm radio continuum observations and {\it Fermi} timing data toward the $\gamma$-ray and radio pulsar PSR\,J0002+6216. We discovered a bow-shock PWN G\,117.33$-$0.07 with a long (4 pc), narrowly collimated non-thermal tail. We estimated the physical properties of G\,117.33$-$0.07 and found that it was strikingly similar to another bow-shock PWN G\,315.78$-$0.23 \citep{nbg+12}. They are both shaped by the large spin-down energy and high (inferred) pulsar velocities, leading to high Mach number shocks with small, compact heads and long synchrotron-emitting tails. In the case of G\,117.33$-$0.07 we have measured the pulsar proper motion and confirmed the high velocity.

 We found that the tail of G\,117.33$-$0.07 points back toward the geometric center of the SNR CTB\,1, suggesting a physical association that is similar to one claimed for the pulsar PSR\,J1437$-$5959, its PWN G\,315.78$-$0.23 and the SNR G\,315.9$-$0.0 \cite[aka The Frying Pan;][]{nbg+12}. Our proper motion measurement is of the right magnitude and direction to support the claim that PSR\,J0002+6216 is a high-velocity pulsar that has escaped its parent SNR CTB\,1. Nonetheless, problems remain and the association, while plausible, is not yet demonstrated. The DM-based distance for \,J0002+6216 of ~7 kpc is inconsistent with the CTB\,1 distance, and its characteristic age of 306 kyr greatly exceeds the SNR age of 10 kyr. In \S\ref{discussion} we have shown how this age and distance for PSR\,J0002+6216 are likely gross overestimates, but it remains the case that a secure PSR-SNR association requires agreement of {\it independently} measured distances and ages. 
 
 A more robust inference, independently arrived at from the pulsar proper motion and the PWN properties, is that PSR\,J0002+6216 is a high-velocity pulsar ($V_{PSR}>$1000 km\,s$^{-1}$). This conclusion does not require the PSR-SNR association. We have adopted a conservative distance of 2 kpc, while a larger distance as inferred from the DM would only serve to increase $V_{PSR}$. We looked at different kick mechanisms for the origin of this very high pulsar velocity. While we lack the data to test whether the rotation axis of the pulsar is aligned with the velocity vector, we note that the large birth period estimated for a high-velocity pulsar like PSR\,J0002+6216 is a general outcome of some impulsive kick models. There is evidence that the heavy elements from CTB\,1 are ejected asymmetrically, in support of the hydrodynamic kick model. However, a reanalysis of the {\it Suzaku} data could test whether the ejecta are preferentially found with direction opposite the pulsar's motion.
 
 The main limitation of this present work is the lack of a well-determined pulsar distance. A radio pulsar parallax measurement would settle the distance debate while also confirming the proper motion obtained by pulsar timing. The analysis of the PWN properties would be substantially improved with more radio data spanning a larger frequency range, and with greater angular resolution. Spectrally resolved X-ray and radio images could be used to help understand why the tail is so uniformly bright over most its length by looking for signatures of in situ particle acceleration of the shocked wind or evidence of synchrotron cooling. Of more immediate importance would be an independent measure of the pulsar velocity and the PWN from deep H$\alpha$ imaging and spectroscopy \citep[e.g.,][]{2010ApJ...724..908R}. In this regard PSR\,J0002+6216 has the same distance and large velocity as PSR\,B2224+65, known for its prominent H$\alpha$ ``Guitar'' nebula \citep{2004ApJ...600L..51C}. Scaling relations from \citet{2014ApJ...784..154B} suggest that the putative bow shock nebula would be compact ($\theta\propto\dot{{E}}^{1/2}/d^2$) but bright ($f_{{\rm H}\alpha}\propto\dot{{E}}$/$d^{7/2}$), provided that the larger $\dot{E}$ of PSR\,J0002+6216  (1.53$\times 10^{35}$ erg s$^{-1}$ vs 1.2 $\times 10^{33}$ erg s$^{-1}$) does not ionize the H\,{\sc i} in the surrounding ISM \citep{lrd82,yuk04}. 

\acknowledgments

The National Radio Astronomy Observatory is a facility of the National Science Foundation operated under cooperative agreement by Associated Universities, Inc. 

The \textit{Fermi} LAT Collaboration acknowledges generous ongoing support
from a number of agencies and institutes that have supported both the
development and the operation of the LAT as well as scientific data analysis.
These include the National Aeronautics and Space Administration and the
Department of Energy in the United States, the Commissariat \`a l'Energie Atomique
and the Centre National de la Recherche Scientifique / Institut National de Physique
Nucl\'eaire et de Physique des Particules in France, the Agenzia Spaziale Italiana
and the Istituto Nazionale di Fisica Nucleare in Italy, the Ministry of Education,
Culture, Sports, Science and Technology (MEXT), High Energy Accelerator Research
Organization (KEK) and Japan Aerospace Exploration Agency (JAXA) in Japan, and
the K.~A.~Wallenberg Foundation, the Swedish Research Council and the
Swedish National Space Board in Sweden.
 
Additional support for science analysis during the operations phase is gratefully
acknowledged from the Istituto Nazionale di Astrofisica in Italy and the Centre
National d'\'Etudes Spatiales in France. This work performed in part under DOE
Contract DE-AC02-76SF00515.

This research has made use the High Energy SNR catalog of \citet{2012AdSpR..49.1313F}\footnote{\url{http://www.physics.umanitoba.ca/snr/SNRcat/}}. This research has made use of the NASA/IPAC Extragalactic Database (NED) which is operated by the Jet Propulsion Laboratory, California Institute of Technology, under contract with the National Aeronautics and Space Administration.  The research presented in this paper has used data from the Canadian Galactic Plane Survey, a Canadian project with international partners, supported by the Natural Sciences and Engineering Resources Council. Work at NRL is supported by NASA. DAF thanks Paul Demorest, Shri Kulkarni, Sterl Phinney and Roger Romani for useful discussions.  MK thanks David Smith for pointing out the connection with HD 225160. We thank the anonymous referee for useful comments that improved the quality of this manuscript.

%
\facilities{VLA, Fermi}

\software{CASA}




{}

\begin{thebibliography}{}

\bibitem[Akaike(1973)]{Akaike73} Akaike, H.\ 1973, 2nd International Symposium on Information Theory, 267

\bibitem[Atwood et al.(2009)]{Atwood09} Atwood, W.~B., Abdo, A.~A., Ackermann, M., et al.\ 2009, \apj, 697, 1071 

\bibitem[Atwood et al.(2013)]{Atwood13} Atwood, W., Albert, A., Baldini, L., et al.\ 2013, arXiv:1303.3514 

\bibitem[Bear \& Soker(2018)]{2018ApJ...855...82B} Bear, E., \& Soker, N.\ 2018, \apj, 855, 82 

\bibitem[Berger(2014)]{berger2014} Berger, E.\ 2014, \araa, 52, 43 

\bibitem[Barkov \& Lyutikov(2018)]{bl18} Barkov, M.~V., \& Lyutikov, M.\ 2018, arXiv:1804.07327 

\bibitem[Bhatnagar et al.(2013)]{brg13} Bhatnagar, S., Rau, U., \& Golap, K.\ 2013, \apj, 770, 91 

\bibitem[Brownsberger \& Romani(2014)]{2014ApJ...784..154B} Brownsberger, S., \& Romani, R.~W.\ 2014, \apj, 784, 154 

\bibitem[Bruel et al.(2018)]{Bruel18} Bruel, P., Burnett, T.~H., Digel, S.~W., et al.\ 2018, arXiv e-prints , arXiv:1810.11394.

\bibitem[Bucciantini(2014)]{2014AN....335..234B} Bucciantini, N.\ 2014, Astronomische Nachrichten, 335, 234 

\bibitem[Buccheri et al.(1983)]{Buccheri83} Buccheri, R., Bennett, K., Bignami, G.~F., et al.\ 1983, \aap, 128, 245 

\bibitem[Bykov et al.(2017)]{bap+17} Bykov, A.~M., Amato, E., Petrov, A.~E., Krassilchtchikov, A.~M., \& Levenfish, K.~P.\ 2017, \ssr, 207, 235 

\bibitem[Camilo et al.(1994)]{ctk94} Camilo, F., Thorsett, S.~E., \& Kulkarni, S.~R.\ 1994, \apjl, 421, L15 

\bibitem[Camilo et al.(2009)]{cng+09} Camilo, F., Ng, C.-Y., Gaensler, B.~M., et al.\ 2009, \apjl, 703, L55 

\bibitem[Chatterjee \& Cordes(2004)]{2004ApJ...600L..51C} Chatterjee, S., \& Cordes, J.~M.\ 2004, \apjl, 600, L51 

\bibitem[Choi et al.(2014)]{2014ApJ...790...99C} Choi, Y.~K., Hachisuka, K., Reid, M.~J., et al.\ 2014, \apj, 790, 99 

\bibitem[Clark et al.(2017)]{cwp+17} Clark, C.~J., Wu, J., Pletsch, H.~J., et al.\ 2017, ApJ, 834, 106

\bibitem[Cordes \& Lazio(2002)]{cl02} Cordes, J.~M., \& Lazio, T.~J.~W.\ 2002, arXiv:astro-ph/0207156 

\bibitem[Craig et al.(1997)]{chp87} Craig, W.~W., Hailey, C.~J., \& Pisarski, R.~L.\ 1997, \apj, 488, 307 

\bibitem[Dickel \& Willis(1980)]{dw80} Dickel, J.~R., \& Willis, A.~G.\ 1980, \aap, 85, 55 

\bibitem[Ferrand \& Safi-Harb (2012)]{2012AdSpR..49.1313F} Ferrand, G., \& Safi-Harb, S.\ 2012, Advances in Space Research, 49, 1313 

\bibitem[Fesen et al.(1997)]{fwr+97} Fesen, R.~A., Winkler, F., Rathore, Y., et al.\ 1997, \aj, 113, 767 

\bibitem[Fich(1986)]{fich86} Fich, M.\ 1986, \apj, 303, 465

\bibitem[Foreman-Mackey et al.(2013)]{emcee} Foreman-Mackey, D., Hogg, D.~W., Lang, D., et al.\ 2013, Publications of the Astronomical Society of the Pacific, 125, 306.

\bibitem[Frail \& Kulkarni(1991)]{fk91} Frail, D.~A., \& Kulkarni, S.~R.\ 1991, \nat, 352, 785 

\bibitem[Frail et al.(1994)]{1994ApJ...437..781F} Frail, D.~A., Goss, W.~M., \& Whiteoak, J.~B.~Z.\ 1994, \apj, 437, 781 

\bibitem[Frail et al.(1996)]{fggd96} Frail, D.~A., Giacani, E.~B., Goss, W.~M., \& Dubner, G.\ 1996, \apjl, 464, L165 

\bibitem[Frail \& Scharringhausen(1997)]{1997ApJ...480..364F} Frail, D.~A., \& Scharringhausen, B.~R.\ 1997, \apj, 480, 364

\bibitem[Fryer et al.(1998)]{1998ApJ...496..333F} Fryer, C., Burrows, A., \& Benz, W.\ 1998, \apj, 496, 333 

\bibitem[Gaensler et al.(2000)]{2000MNRAS.318...58G} Gaensler, B.~M., Stappers, B.~W., Frail, D.~A., et al.\ 2000, \mnras, 318, 58 

\bibitem[Gaia Collaboration et al.(2018)]{2018A&A...616A...1G} Gaia Collaboration, Brown, A.~G.~A., Vallenari, A., et al.\ 2018, \aap, 616, A1 
\bibitem[Gvaramadze(2004)]{gva04} Gvaramadze, V.~V.\ 2004, \aap, 415, 1073 
\bibitem[Haffner et al.(2003)]{2003ApJS..149..405H} Haffner, L.~M., Reynolds, R.~J., Tufte, S.~L., et al.\ 2003, \apjs, 149, 405 

\bibitem[Hailey \& Craig(1994)]{hc94} Hailey, C.~J., \& Craig, W.~W.\ 1994, \apj, 434, 635 

\bibitem[Helfand et al.(2001)]{2001ApJ...556..380H} Helfand, D.~J., Gotthelf, E.~V., \& Halpern, J.~P.\ 2001, \apj, 556, 380 

\bibitem[Hester \& Kulkarni(1988)]{1988ApJ...331L.121H} Hester, J.~J., \& Kulkarni, S.~R.\ 1988, \apjl, 331, L121 

\bibitem[Hobbs et al.(2005)]{2005MNRAS.360..974H} Hobbs, G., Lorimer, D.~R., Lyne, A.~G., \& Kramer, M.\ 2005, \mnras, 360, 974 

\bibitem[Holland-Ashford et al.(2017)]{hla+17} Holland-Ashford, T., Lopez, L.~A., Auchettl, K., Temim, T., \& Ramirez-Ruiz, E.\ 2017, \apj, 844, 84 

\bibitem[Iben \& Tutukov(1996)]{1996ApJ...456..738I} Iben, I., Jr., \& Tutukov, A.~V.\ 1996, \apj, 456, 738 

\bibitem[Intema et al.(2017)]{ijmf17}  Intema, H.~T., Jagannathan, P., Mooley, K.~P., \& Frail, D.~A.\ 2017, \aap, 598, A78

\bibitem[Janka(2017)]{janka2017} Janka, H.-T.\ 2017, \apj, 837, 84 

\bibitem[Johnston et al.(2007)]{2007MNRAS.381.1625J} Johnston, S., Kramer, M., Karastergiou, A., et al.\ 2007, \mnras, 381, 1625 


\bibitem[Kargaltsev et al.(2008)]{2008ApJ...684..542K} Kargaltsev, O., Misanovic, Z., Pavlov, G.~G., Wong, J.~A., \& Garmire, G.~P.\ 2008, \apj, 684, 542 

\bibitem[Kargaltsev et al.(2017)]{2017JPlPh..83e6301K} Kargaltsev, O., Pavlov, G.~G., Klingler, N., \& Rangelov, B.\ 2017, Journal of Plasma Physics, 83, 635830501 

\bibitem[Kaspi(1996)]{kaspi96} Kaspi, V.~M.\ 1996, IAU Colloq.~160: Pulsars: Problems and Progress, 105, 375 

\bibitem[Katsuda et al.(2018)]{kmj+18} Katsuda, S., Morii, M., Janka, H.-T., et al.\ 2018, \apj, 856, 18 

\bibitem[Katsuragawa et al.(2018)]{knm+18} Katsuragawa, M., Nakashima, S., Matsumura, H., et al.\ 2018, \pasj,  

\bibitem[Kent et al.(2018)]{2018AAS...23134214K} Kent, B.~R., Masters, J.~S., Chandler, C.~J., et al.\ 2018, American Astronomical Society Meeting Abstracts \#231, 231, 342.14 

\bibitem[Kerr(2011)]{Kerr11} Kerr, M.\ 2011, \apj, 732, 38 

\bibitem[Kerr et al.(2015)]{Kerr15} Kerr, M., Ray, P.~S., Johnston, S., Shannon, R.~M., \& Camilo, F.\ 2015, \apj, 814, 128 


\bibitem[Kochanek et al.(2018)]{2018arXiv181008620K} Kochanek, C.~S., Auchettl, K., \& Belczynski, C.\ 2018, arXiv:1810.08620 

\bibitem[Kothes et al.(2006)]{kffu06} Kothes, R., Fedotov, K., Foster, T.~J., \& Uyan{\i}ker, B.\ 2006, \aap, 457, 1081 

\bibitem[Kothes(2017)]{kothes17} Kothes, R.\ 2017, Modelling Pulsar Wind Nebulae, Astrophysics and Space Science Library, 446, 1 

\bibitem[Kulkarni et al.(1993)]{kpha93} Kulkarni, S.~R., Predehl, P., Hasinger, G., \& Aschenbach, B.\ 1993, \nat, 362, 135 

\bibitem[Kulkarni et al.(2014)]{2014ApJ...797...70K} Kulkarni, S.~R., Ofek, E.~O., Neill, J.~D., Zheng, Z., \& Juric, M.\ 2014, \apj, 797, 70 

\bibitem[Lai et al.(2001)]{2001ApJ...549.1111L} Lai, D., Chernoff, D.~F., \& Cordes, J.~M.\ 2001, \apj, 549, 1111 

\bibitem[Landecker et al.(1982)]{lrd82} Landecker, T.~L., Roger, R.~S., \& Dewdney, P.~E.\ 1982, \aj, 87, 1379 

\bibitem[Lazendic \& Slane(2006)]{ls06} Lazendic, J.~S., \& Slane, P.~O.\ 2006, \apj, 647, 350 

\bibitem[Lopez \& Fesen(2018)]{lf18} Lopez, L.~A., \& Fesen, R.~A.\ 2018, \ssr, 214, 44 

\bibitem[Manchester et al.(2005)]{2005AJ....129.1993M} Manchester, R.~N., Hobbs, G.~B., Teoh, A., \& Hobbs, M.\ 2005, \aj, 129, 1993 

\bibitem[McMullin et al.(2007)]{2007ASPC..376..127M} McMullin, J.~P., Waters, B., Schiebel, D., Young, W., \& Golap, K.\ 2007, Astronomical Data Analysis Software and Systems XVI, 376, 127 

\bibitem[Ng \& Romani(2004)]{2004ApJ...601..479N} Ng, C.-Y., \& Romani, R.~W.\ 2004, \apj, 601, 479 

\bibitem[Ng \& Romani(2007)]{2007ApJ...660.1357N} Ng, C.-Y., \& Romani, R.~W.\ 2007, \apj, 660, 1357 
\bibitem[Ng et al.(2012)]{nbg+12} Ng, C.-Y., Bucciantini, N., Gaensler, B.~M., et al.\ 2012, \apj, 746, 105 

\bibitem[Nicastro et al.(1996)]{njk96} Nicastro, L., Johnston, S., \& Koribalski, B.\ 1996, \aap, 306, L49 

\bibitem[Pacholczyk(1970)]{1970ranp.book.....P} Pacholczyk, A.~G.\ 1970, Series of Books in Astronomy and Astrophysics, San Francisco: Freeman, 1970, 

\bibitem[Panagia(1973)]{1973AJ.....78..929P} Panagia, N.\ 1973, \aj, 78, 929 

\bibitem[Pannuti et al.(2010)]{prb+10} Pannuti, T.~G., Rho, J., Borkowski, K.~J., \& Cameron, P.~B.\ 2010, \aj, 140, 1787 

\bibitem[Perley et al.(2009)]{2009IEEEP..97.1448P} Perley, R., Napier, P., Jackson, J., et al.\ 2009, IEEE Proceedings, 97, 1448 

\bibitem[Popov \& Turolla(2012)]{pt12} Popov, S.~B., \& Turolla, R.\ 2012, \apss, 341, 457 

\bibitem[Prentice \& Ter Haar(1969)]{1969MNRAS.146..423P} Prentice, A.~J.~R., \& Ter Haar, D.\ 1969, \mnras, 146, 423 

\bibitem[Rau \& Cornwell(2011)]{rc11} Rau, U., \& Cornwell, T.~J.\ 2011, \aap, 532, A71 

\bibitem[Rau et al.(2016)]{rbo16} Rau, U., Bhatnagar, S., \& Owen, F.~N.\ 2016, AJ, 152, 124

\bibitem[Reid et al.(2014)]{2014ApJ...783..130R} Reid, M.~J., Menten, K.~M., Brunthaler, A., et al.\ 2014, \apj, 783, 130 

\bibitem[Rengelink et al.(1997)]{rtb+97} Rengelink, R.~B., Tang, Y., de Bruyn, A.~G., et al.\ 1997, \aaps, 124,  259

\bibitem[Romani et al.(2010)]{2010ApJ...724..908R} Romani, R.~W., Shaw, M.~S., Camilo, F., Cotter, G., \& Sivakoff, G.~R.\ 2010, \apj, 724, 908 

\bibitem[Sakai et al.(2018)]{2018IAUS..336..168S} Sakai, N., \& BeSSeL and VERA Projects Members 2018, Astrophysical Masers: Unlocking the Mysteries of the Universe, 336, 168 

\bibitem[Sale et al.(2014)]{sale14} Sale, S.~E., Drew, J.~E., Barentsen, G., et al.\ 2014, \mnras, 443, 2907 

\bibitem[Scheck et al.(2006)]{2006A&A...457..963S} Scheck, L., Kifonidis, K., Janka, H.-T., \& M{\"u}ller, E.\ 2006, \aap, 457, 963 

\bibitem[Shternin et al.(2017)]{2017JPhCS.932a2004S} Shternin, P.~S., Yu, M., Kirichenko, A.~Y., et al.\ 2017, Journal of Physics Conference Series, 932, 012004 

\bibitem[Slane(2017)]{2017hsn..book.2159S} Slane, P.\ 2017, Handbook of Supernovae, ISBN 978-3-319-21845-8.~Springer International Publishing AG, 2017, p.~2159, 2159 

\bibitem[Spruit \& Phinney(1998)]{sp98} Spruit, H., \& Phinney, E.~S.\ 1998, \nat, 393, 139 

\bibitem[Toropina et al.(2018)]{trl18} Toropina, O.~D., Romanova, M.~M., \& Lovelace, R.~V.~E.\ 2018, arXiv:1803.06240 

\bibitem[van der Swaluw et al.(2003)]{vag+03} van der Swaluw, E., Achterberg, A., Gallant, Y.~A., Downes, T.~P., \& Keppens, R.\ 2003, \aap, 397, 913 

\bibitem[Verbunt et al.(2017)]{vi2017} Verbunt, F., Igoshev, A., \& Cator, E.\ 2017, \aap, 608, A57 

\bibitem[Vigna-G{\'o}mez et al.(2018)]{2018MNRAS.481.4009V} Vigna-G{\'o}mez, A., Neijssel, C.~J., Stevenson, S., et al.\ 2018, \mnras, 481, 4009 

\bibitem[Wenger et al.(2018)]{2018ApJ...856...52W} Wenger, T.~V., Balser, D.~S., Anderson, L.~D., \& Bania, T.~M.\ 2018, \apj, 856, 52 
\bibitem[Wilks(1938)]{Wilks38} Wilks, S.~S.\ 1938, Ann. Math. Statist., 9, 60

\bibitem[Wongwathanarat et al.(2013)]{2013A&A...552A.126W} Wongwathanarat, A., Janka, H.-T., \& M{\"u}ller, E.\ 2013, \aap, 552, A126 

\bibitem[Wu et al.(2018)]{wu+18} Wu, J., Clark, C.~J., Pletsch, H.~J., et al.\ 2018, \apj, 854, 99 

\bibitem[Yao et al.(2017)]{ymw17} Yao, J.~M., Manchester, R.~N., \& Wang, N.\ 2017, \apj, 835, 29 
\bibitem[Yar-Uyaniker et al.(2004)]{yuk04} Yar-Uyaniker, A., Uyaniker, B., \& Kothes, R.\ 2004, \apj, 616, 247 
\bibitem[Zeiger et al.(2008)]{2008ApJ...674..271Z} Zeiger, B.~R., Brisken, W.~F., Chatterjee, S., \& Goss, W.~M.\ 2008, \apj, 674, 271 

\bibitem[Zyuzin et al.(2018)]{zks18} Zyuzin, D.~A., Karpova, A.~V., \& Shibanov, Y.~A.\ 2018, \mnras, 476, 2177 




\end{thebibliography}
\end{document}